\documentclass[twocolumn,aps,showpacs]{revtex4}
\usepackage{dcolumn}
\usepackage{bm}
\usepackage{epsfig}
\usepackage{epsf}

\begin{document}
\title{ Imaging Studies of Characteristics for a Slab of a Lossy Left-handed Material}
\author{Linfang Shen$^{1}$ and Sailing He$^{1,2}$}
\address{$^{1}$
Centre for Optical and Electromagnetic Research,\\
State Key Laboratory for Modern Optical Instrumentation, \\
Zhejiang University, \\
Yu-Quan, Hangzhou 310027, P. R. China\\
$^{2}$Division of Electromagnetic Theory, Alfven Laboratory, Royal Institute of Technology, \\
S-100 44 Stockholm, Sweden
}
\date{\today}

\begin{abstract}
The characteristics of an imaging system formed by a slab of a lossy left-handed material (LHM) are studied. The transfer
function of the LHM imaging system is written in an appropriate product form with each term having a clear physical interpretation. A tiny loss of the LHM may suppress the transmission of evanescent waves through the LHM slab and this is explained physically.
An analytical expression for the resolution of the imaging
system is derived. It is shown that it is impossible to make a subwavelength imaging by using a realistic LHM imaging system unless the LHM slab is much thinner than the wavelength.

\end{abstract}
\maketitle
\section{Introduction}

Materials that have simultaneously a negative electric permittivity and a negative magnetic
permeability have attracted much attention recently \cite{r1,r2,r3,r4}. These materials have extraordinary properties, such as the negative refraction, inverse Doppler shift and  backward-directed Cherenkov radiation cone, etc \cite{veselago}. The electric field, the magnetic field and the wave vector of an electromagnetic wave propagating in such a material obey the left-hand rule and thus it is called the left-handed material (LHM). Due to its effect of the negative refraction, a LHM planar slab can act as a lens, and focusing waves from a point source \cite{veselago}. Recently, Pendry predicted that a LHM slab can amplify evanescent waves and thus behaves like a perfect lens \cite{pendry}. It is well-known that a conventional imaging system is limited in resolution by
the wavelength of the used light, since the information finer than the wavelength is
carried by evanescent waves, which are drastically decayed before they reach the image plane. A LHM slab appears attractive in provideing an effective approach to achieve a subwavelength imaging. However, the LHM that Pendry considered is quite ideal and has no loss. Garcia et al \cite{garcia} pointed out later that an actual LHM should be lossy and the LHM loss may completely suppress the amplification of evanescent waves, and consequently a LHM slab can not act as a perfect lens. It is interesting to find out answers to the following questions: (1) What is the characteristic of a realistic LHM imaging system? (2) How does the resolution of the system depend on the LHM loss and other slab parameters?
(3) Is it possible to use a realistic LHM imaging system to make a subwavelength imaging? If yes, then how to realize it?  In the present paper,  we will study these problems theoretically.

\section{ Transfer function of a LHM imaging system}

Consider a LHM slab imaging system with co-ordinates as shown in Fig. 1.
We assume that the system is invariant along the y-axis direction and
only the one-dimensional imaging will be considered here for simplicity. In the imaging system,
the space can be divided into three regions: region 1 ($z<{0}$), where the objective plane is located; region 2 (the LHM slab); region 3 ($z\geq{d}$), where the image plane is located. The electric permittivity and magnetic permeability in the three regions are denoted
by $\epsilon_l$ and $\mu_l$ ($l=1,2,3$), respectively. Here $\epsilon_1=\epsilon_3\ge{0}$, $\mu_1=\mu_3\ge{0}$, $Re\epsilon_2<{0}$, and $Re\mu_2<{0}$.
This imaging system has the properties of being linear and invariant under
translation along the $x$ direction, and thus can be described by a transfer function
(${\cal H}$) or impulse
response ($h$) (the latter is just the Fourier transform of the former).

In order to calculate the transfer function of the LHM imaging system, we consider a plane wave
propagating along the z direction from the objective plane $z=-a$ ($a > 0$). The plane wave will refract and reflect simultaneously at the interfaces $z=0$ and $z=d$ before it reaches the image plane $z=d+b$ (at a distance of b away from the LHM slab). The electromagnetic fields can be decomposed into the $E$- and $H$-polarization modes. Let's first consider the $E$-polarization case, where the electric field has only $y$ component (i.e., $\vec{E}=E\hat{y}$), and the magnetic field has the form of $\vec{H}=H_x\hat{x}+H_z\hat{z}$. All the field components are assumed to have the time dependence of form $\exp ( - i\omega t)$. The electric field $E$ of the wave in the three regions can be written in the following form of
\begin{eqnarray}
E_1 (x,z) &=& \exp({ik_xx})\bigl\{A_1\exp({ik_z(z+a)})+B_1\exp({-ik_zz})\bigr\},\eqnum{1.1}\\
E_2 (x,z) &=& \exp({ik_xx})\bigl\{A_2\exp({-iqz})+B_2\exp({iqz})\bigr\},\eqnum{1,2}\\
E_3 (x,z) &=& A_3\exp({ik_xx})\cdot\exp(ik_z(z-d)),\eqnum{1,3}
\end{eqnarray}
where $k_x$ is the $x$-component of the wave vector, $k_z$ is the z-component in regions 1 and 3, and $q$ is the $z$-component of the wave vector in region 2.   Obviously, one has $k_z=\sqrt{{n_1^2}k_0^2-k_x^2}$ for propagating waves (i.e., $k_x < n_1 k_0$, where $n_1=\sqrt{\epsilon_1\mu_1}$), and $k_z=i\sqrt{k_x^2-{n_1^2}k_0^2}$ for evanescent waves (i.e., $k_x \ge n_1 k_0$). Similarly, $q=\sqrt{\epsilon_2 \mu_2 k_0^2-k_x^2}$ for propagating waves and $q=i\sqrt{k_x^2-\epsilon_2 \mu_2 k_0^2}$ for evanescent waves. The coefficients for the forward and backward (concerning the energy flow) propagating components are denoted by $A$ and $B$, respectively. Note that the energy flow for the term $ A_2\exp({-iqz})$ in Eq.(1.2) is along the $+z$ direction since the refraction index of the LHM slab is negative.

The transfer function of the imaging system is defined as ${\cal H}(k_x)=E_3(x,d+b)/A_1\exp(ik_xx)$, i.e., ${\cal H}(k_x)=(A_3 /A_1 )\exp (ik_z b)$. From Maxwell's equations, one has $H_x=(i/\omega\mu_0\mu)(\partial{E}/\partial{z})$. From the continuity condition of the field components of $E$ and $H_x$ at the interfaces $z=0$ and $z=d$, one can derive formulas for $A_3/A_1$ and ${\cal H}(k_x)$. In the present paper we can write the transfer function ${\cal H}(k_x)$ in the following product form (with each term having a clear physical interpretation)
\begin{eqnarray}
{\cal H}(k_x) &=& t_1Pt_2\bar{P}, \eqnum{2} \label{e2}
\end{eqnarray}
where
\begin{eqnarray}
t_1 &=& {2\over{(1-\xi)+(1+{\xi})P^2}}{1\over{t_2}}, \eqnum{3.1} \label{e3.1}\\
t_2 &=& 2(1-{{\mu_2k_z}\over{\mu_1q}})^{-1}, \eqnum{3.2} \label{e3.2}\\
P &=& \exp(-iqd), \eqnum{3.3} \label{e3.3}\\
\bar{P} &=& \exp\{ik_z(a+b)\},\eqnum{3.4} \label{e3.4}\\
\xi &=& {1\over{2}}\Bigl({{\mu_1q}\over{\mu_2k_z}}+{{\mu_2k_z}\over{\mu_1q}}\Bigr).
\eqnum{3.5}\label{e3.5}
\end{eqnarray}
Here $t_1$ represents the effective transmission coefficient for a plane wave at the interface $z=0$ (i.e., $t_1=A_2/(A_1exp(ik_za))$), while $t_2$ is the simple transmission coefficient at $z=d$ (i.e., $t_2=A_3/(A_2exp(-iqd))$).  $\xi$ is related to the impedance matching of the LHM  with the surrounding medium ($\xi = -1$ when the impedance is matched).  The second term $(1+{\xi})P^2$ in the denominator of expression (3.1) for $t_1$ is attributed to the
effect of the reflection of the wave at the interface $z=d$.  In fact, this term  ($(1+{\xi})P^2$) vanishes if the LHM slab is completely matched in the impedance with the medium in region 3  (i.e. $\epsilon_2=-\epsilon_3$,
$\mu_2=-\mu_3$) regardless what the value of $\epsilon_1$ or $\mu_1$ is.
$P$ is the phase change factor or amplitude amplification factor as the wave passes the LHM slab, and $\bar{P}$ is the total phase change factor or amplitude decaying factor as the wave goes from the objective plane to the left surface of the LHM slab and from the right surface of the LHM slab to the image plane.

For an ideal LHM imaging system (i.e., $\epsilon_2=-\epsilon_1$, $\mu_2=-\mu_1$ and $b=d-a$), one has $\xi=-1$, $t_1=1$, $t_2=1$, and ${\cal H}(k_x)=1$ for all $k_x$, this  means the imaging system is perfect.

\section{Influence of the loss in the LHM}

It is known that an actual LHM should be lossy (i.e. $\epsilon_2$ or $\mu_2$ has an imaginary part) since any LHM must be dispersive \cite{smith}. Here we analyze the characteristics of the transfer function for an imaging system formed by a lossy LHM slab.
 Consider first the case when the loss comes only from the imaginary part of $\epsilon_2$, i.e.,
\[
\epsilon_2=-\epsilon_1(1-i\delta), \qquad \mu_2=-\mu_1,
\]
where we assume $0<\delta\ll{1}$.
As ${\cal H}$ is an even function of $k_x$, we only needs to consider the interval $[0,\infty)$ for $k_x$. For the convenience of
analysis,  we further divide this interval into two regions, namely, $k_x \in [0, n_1k_0)$ and $k_x \in [n_1k_0, + \infty)$.

1. The case when $k_x < n_1k_0$. This case corresponds to propagating waves with $k_z=\sqrt{{n_1^2}k_0^2-k_x^2}$ being a real number. Then one has $q=\sqrt{\epsilon_2 \mu_2 k_0^2-k_x^2}=\sqrt{k_z^2-i{n_1^2}{k_0^2}\delta}$.

(i) Consider the special situation when the plane waves are propagating nearly along the z axis,  i.e., $k_x \ll\sqrt{1-\delta}n_1 k_0$ (or equivalently, $k_z\gg\sqrt{\delta}{n_1}k_0$). In this case, one has $\xi\cong-1+O(\delta^2)$, which implies that the media are approximately matched and $t_1\approx 1/t_2\approx{1}$. The transfer function can then be approximated by
\begin{eqnarray}
{\cal H}\left( {k_x } \right) \approx \exp \left( { - \frac{{n_1 k_0 }}{{2\sqrt {n_1^2 k_0^2  - k_x^2 } }}D\delta } \right), \eqnum{4} \label{e4}
\end{eqnarray}
where $D=n_1k_0d$. At the point $k_x=0$ (corresponding to the plane wave normally incident on the LHM slab), one has ${\cal H}(0)=\exp(-n_1 k_0 d \delta /2)$.
This indicates that the small loss of the LHM makes $H(0)$ a bit below $1$.

(ii) Then consider the other situation when $k_x\to\ n_1k_0$ (i.e.,
$k_z \to  0$). In this situation, one finds that $k_z \ll \sqrt{\delta}n_1k_0$ and $\xi=-{\sqrt{-i\delta}}{{n_1k_0}/{(2k_z)}}$ differs from $-1$ completely.
Therefore, a small loss of the LHM causes a serious mismatch problem for these waves.
The transfer function can be approximated by
\begin{eqnarray}
{\cal H}(k_x) \approx {{2\sqrt{{n_1^2}k_0^2-k_x^2}}\over{n_1 k_0 D\delta}}\exp\Bigl\{{-D{\sqrt{{\delta}\over{2}}}(1+i)}\Bigr\}, \eqnum{5} \label{e6}
\end{eqnarray}
Obviously, ${\cal H}(k_x)\to0$ when $k_x\to n_1k_0$. Therefore, the existence of a small loss in the LHM changes the value of the ${\cal H}$ function completely (dropping to 0 from 1; see Fig. 2(c) below) in the neighborhood of
$k_x = n_1k_0$.

2. The case when $k_x \ge n_1 k_0$.
This case corresponds to evanescent waves with $k_z=i\sqrt{k_x^2-n_1^2k_0^2}$. Then one has
$q=i\sqrt{k_x^2-\epsilon_2\mu_2k_0^2}=i\sqrt{|k_z|^2+i\delta n_1^2k_0^2}$.

(i)  Consider first the situation when
$k_x\to n_1k_0$ (i.e., $k_z \to  0$). In this situation, one finds that $|k_z|\ll\sqrt{\delta}n_1k_0$
and $\xi\approx-{{\sqrt{i\delta}}n_1k_0/({2|k_z|})}$ differs very much from $-1$. Thus, one sees that a small loss of the LHM causes a serious mismatch problem for these evanescent waves. One can then approximate
the transfer function as
\begin{eqnarray}
{\cal H}(k_x) \approx i{{2\sqrt{k_x^2-n_1 k_0^2}}\over{n_1 k_0 D\delta}}\exp\Bigl\{{D{\sqrt{{\delta}\over{2}}}(1+i)}\Bigr\}. \eqnum{6} \label{e7}
\end{eqnarray}
Obviously, ${\cal H}\to 0$ when $k_x\to n_1 k_0$. We find that the real part (${\cal H}_r$) of
${\cal H}$ is far smaller than the imaginary part (${\cal H}_i$) in this case. ${\cal H}_i$ decreases as $k_x $ decreases to $n_1 k_0$. On the other hand,  ${\cal H}_i$ should approach zero when $k_x\gg n_1 k_0$ \cite{garcia}.  Thus, there exists a peak of
${\cal H}_i$ on the right-hand side of the point $k_x=n_1 k_0$ (see Fig. 2(d) below).

(ii) Then consider the other situation when
$k_x\gg\sqrt{1+\delta}n_1k_0$ (or equivalently, $|k_z|\gg\sqrt{\delta}n_1k_0$). These  evanescent waves are very important for a subwavelength imaging. For this case our analysis shows that $\xi=-1+(n_1 k_0/k_z)^4\delta^2/8$, which indicates that $\delta$ causes only a second-order perturbation to $\xi$ (the LHM slab seems still well-matched with the surrounding media for these evanescent waves). However, $|P|\approx exp(\sqrt{k_x^2-n_1^2 k_0^2}d)$ increases exponentially as $k_x$ increases. Thus, unlike the case of propagating waves, the second term $(1+\xi)P^2$ in the denominator of the expression (3.1) for $t_1$
can not be neglected. In other words, a tiny $\delta$ may cause a significant change of the effective transmission coefficient $t_1$ for the evanescent wave at the interface $z=0$  (note that $\delta$ causes only a small perturbation to $t_2$ and
$P$), and consequently leads to a quick decrease of $|{\cal H}|$.
For a slab of usual medium (with or without loss), the effective transmission coefficient $t_1$  must be close to 1 if the impedance is almost matched
(regardless the value of $k_x$). However, for a LHM slab $t_1$ can be very small (consequently the total transmission coefficient of the LHM slab can  be very small) even the
the impedance is almost matched. The main reason is that
the evanescent wave "reflected" at the back interface of the LHM slab may be effectively amplified (before it comes
out of the slab) and thus significantly suppress the transmission of the evanescent wave according to Eq. (3.1). Therefore,
the conventional concept of impedance match  is insufficient to describe the medium match for evanescent waves for a LHM slab.
We introduce a critical value ($k_x^c$) for $k_x$, at which $Re{{\cal H}}\approx1/2$. Apparently, $k_x^c$ is closely related to the resolution of the imaging system. After a perturbation analysis for Eqs. (2)-(3.5), one can conclude that the critical value $k_x^c$ can be approximately determined from the condition $ Re[(\xi+1)P^2] = 2$ (physically this condition can be understood as a balance condition for the two terms in the denominator of expression (3.1) for $t_1$).
When $k_x\gg k_x^c$, the term $(\xi+1)P^2$ becomes dominant in the denominator of expression (3.1) for $t_1$, and $t_1$ decreases rapidly to 0 (consequently ${\cal H}\to0$) as $k_x$ increases. The transfer function can then be approximated as
\begin{eqnarray}
{\cal H}(k_x) \approx 16\Bigl({{k_x^2-n_1^2 k_0^2}\over{n_1^2 k_0^2}\delta}\Bigr)^2\exp({-2\sqrt{k_x^2-n_1^2 k_0^2}d}), \qquad k_x\gg k_x^c. \eqnum{7} \label{e8}
\end{eqnarray}

Figs. 2(a) and 2(b) show the real and imaginary parts of ${\cal H}(k_x)$ (calculated with Eq. (2)), respectively, where we have chosen $d=\lambda$ ($\lambda=2\pi/n_1k_0$) and $\delta=10^{-4}$. Figs. 2(c) and 2(d) give the corresponding enlarged views of ${\cal H}_r(k_x)$ and ${\cal H}_i(k_x)$ in the neighborhood of $n_1k_0$. From Figs. 2(b)and 2(d) one sees that ${\cal H}_i$ has a peak around $ k_x =n_1k_0$ (a bit larger than $ n_1k_0$) as discussed before. Nevertheless, ${\cal H}_i$ can be approximated to zero everywhere (except in a very narrow
region around $n_1k_0$; this non-zero region is so narrow that it can be neglected in an imaging system for the image in the physical space). Our numerical analysis shows
that $k_x^c\approx2.3n_1k_0$ for this example.
Fig. 2(a) shows that ${\cal H}$ decreases to zero rapidly as $k_x$ increases in the region $k_x\ge k_x^c$. From these figures one sees that the characteristics of the transfer function ${\cal H}$ (calculated from Eq. (2)) agree very well with our earlier analysis.

The value of $k_x^c$ is very important for an imaging system  since $k_x^c$ directly relates the smallest scale in which information can be correctly transferred by the system. To see clearly what influences this parameter, we wish to derive an analytic formula for $k_x^c$. From $ Re[ (\xi+1)P^2] = 2$, one obtains $\gamma e^{- \gamma}= p$ (i.e.,
$\ln \gamma - \gamma = \ln p$),
where $\gamma=D\sqrt{(k_x^c)^2-n_1^2k_0^2}/(2n_1 k_0)$ and $p= D \sqrt{\delta}/4 $. We assume that loss parameter $\delta $ is so small that $ p \ll 1$.
Then one has $e^{\gamma} \gg \gamma >1$, i.e., $ \gamma \gg \ln \gamma >0$.
Thus it follows that $ \gamma = - \ln p + \ln \gamma \approx - \ln p $. This approximation can be improved
as $ \gamma \approx - \ln p + \ln \ln (1/p)$.
One can finally obtains the following analytic formula
\begin{eqnarray}
k_x^c  &\approx& \left\{ {\left[ {\frac{2}{D}\ln \left( {\frac{4}{{D\sqrt \delta  }}} \right) + \frac{2}{D}\ln \ln \left( {\frac{4}{{D\sqrt \delta  }}} \right)} \right]^2  + 1} \right\}^{\frac{1}{2}} \left( {n_1 k_0 } \right)  \eqnum{8}\label{e8}.
\end{eqnarray}
Note that $D=n_1 k_0 d$.

Eq. (8) shows that for a given loss parameter$\delta$, $k_x^c$ decreases as the thickness of the LHM slab increases (actually $k_x^c$  is almost inverse proportional to $d$ as one can see from $ \gamma \approx - \ln p $). On the other hand, $k_x^c$ is almost linearly proportional to $\ln{\delta}$ (since $\left( {2/D} \right)\ln \left[ {4/\left( {D\sqrt \delta  } \right)} \right] =  - \ln \left( \delta  \right)/D + 2\ln \left( {16/D} \right)/D$ and the first term is dominant over the second term). Thus,
$k_x^c$ does not increase noticeably even if $\delta$ decrease by one order of magnitude.
To realize the subwavelength imaging, e.g., $k_x \ge 10k_0$, it is required that $\delta < 10^{-52}$ when $d=2\lambda$, and $\delta < 2\times10^{-25}$ when $d=\lambda$.
It is doubtful whether the loss of a realistic LHM could be reduced to such a level. Therefore, the thickness of the LHM slab should be smaller than the wavelength in order to realize the subwavelength imaging for such a system.

From Eq. (2)-(3.5) (or Eq. (8)) one can find that the characteristics of the transfer function (as well as  $k_x^c/(n_1k_0)$) depend only on $\delta$ and $d$/$\lambda$. Fig. 3(a) gives the contour plot of $k_x^c/(n_1k_0)$ (calculated with Eq. (2)) as a function of $\delta$ and $d$/$\lambda$. From this figure one sees that $k_x^c$ is very sensitive to the thickness $d$. Our numerical calculation shows that the requirement for $k_x^c =10 n_1k_0$ (for a subwavelength imaging) is $\delta=2\times10^{-25}$  when $d=\lambda$,
$\delta=9 \times10^{-11}$ when $d=\lambda /2$, and $\delta=6\times10^{-5}$ when $d=\lambda/4$. Fig. 3(b) shows $k_x^c$ obtained from Eq.(2) and the analytic expression (8), respectively, as the loss parameter varies. One sees that they agree very well.

Consider now the impulse response for a lossy LHM imaging system. The impulse response $h(x)$ is defined as  the field distribution on the image plane produced by a point source on the objective plane. Obviously, it  is just the Fourier transform of the transfer function ${\cal H}( k_x )$.
Our analysis shows that ${\cal H}( k_x )$ is approximately a step function, i.e., ${\cal H}(k_x)\approx 1$ for $|k_x|< k_x^c$ (except the abrupt change in the narrow neighborhood of $k_x=n_1 k_0$, which is narrow enough to be neglected in calculating the impulse response by the Fourier transform), while ${\cal H}(k_x)\approx 0$ elsewhere. Thus, we have approximately $h(x)=2k_x^c sin({k_x^c}x)/({k_x^c}x)$. According to the definition of the resolution (denoted by $\Delta$) for an imaging system, we find that $\Delta=\pi/k_x^c$.
Figs. 4(a) and 4(b) show the real and imaginary parts of the impulse response (normalized by $h(0)$) of a LHM imaging system, respectively,  with $\delta=10^{-4}$ and $d=\lambda$. One sees that the imaginary part of the impulse response is very small and can be neglected as compared with the real part. From Fig. 4(a) one sees that the real part of $h$ calculated from the Fourier transform of Eq.(2) (the solid line) agrees very well with that from our analytic expression (the dotted line) in the region of $|x|< 2\pi/k_x^c$ (note that our analytic expression gives a real value of $h$; the imaginary part of $h$ is always very small, as show in Fig. 4(b)). This indicates that the resolution expression
$\Delta=\pi/k_x^c$ is quite reliable. Thus, one can use the following explicit resolution for a lossy LHM imaging system
\begin{eqnarray}
\Delta  \approx \left\{ {\left[ {\frac{2}{D}\ln \left( {\frac{4}{{D\sqrt \delta  }}} \right) + \frac{2}{D}\ln \ln \left( {\frac{4}{{D\sqrt \delta  }}} \right)} \right]^2  + 1} \right\}^{ - \frac{1}{2}} \left( {\frac{\pi }{{n_1 k_0 }}} \right). \eqnum{9}\label{e9}
\end{eqnarray}

In the above analysis, we have assumed that the loss of the LHM comes from the imaginary part
of $\epsilon_2$. We can also discuss the case in which $\mu _2$ gives rise to the loss, i.e. $\epsilon_2=-\epsilon_1$, $\mu_2=-\mu_1(1-i\delta)$.
Our analysis and numerical calculation have shown that the characteristics of the transfer function for this case is
similar to the previous case that we have discussed. In this case one has $k_x^c  \approx \left( {n_1 k_0 } \right)\sqrt {\left[ {\left( {1/D} \right)\ln \left( {2/\delta } \right)} \right]^2  + 1}$ and $\Delta  \approx  \left( {\pi /n_1 k_0 } \right)/\sqrt {\left[ {\left( {1/D} \right)\ln \left( {2/\delta } \right)} \right]^2  + 1}$. For the same $d$ and $\delta$, $k_x^c$ for this case is  a bit
smaller than that for the previous case (see Fig.5).

In the above analysis the electromagnetic wave is assumed to be $E-$ polarized. For  the $H-$polarized case, one can obtain the transfer function from Eq.(2) directly by interchanging $\epsilon$ and $\mu$ due to the symmetry of Maxwell's equations. Thus, similar analytic formulas for $k_x^c$ and $\Delta $ can be found.

\section{Conclusion and discussion}

In this paper we have analyzed the characteristics of the transfer function for  an imaging system formed by a lossy
LHM slab. The small loss in the LHM may completely change the  transfer function in the following aspects: (1) the transfer function drops to zero abruptly near the turning point $k_x = n_1k_0$ for the propagating and
 evanescent waves;
(2) ${\cal H}$ decreases to zero rapidly when $k_x$ increases from a critical $k_x^c$ (at which ${\cal H}={1/2}$). Our analysis shows that $k_x^c$ is almost linearly proportional to the logarithm of the loss parameter $\delta $ (related to the imaginary part of $\epsilon_2$ or $\mu_2$) of the LHM. Thus, $k_x^c$ does not increase noticeably even if the loss decreases by one order of
magnitude. Analytical formulas for $k_x^c$ and the resolution of the imaging
system have been derived. It has been shown that the thickness of the LHM slab should be smaller than the wavelength in order to realize the subwavelength imaging for such a system.

Our analysis has also shown that a tiny loss of the LHM may suppress the transmission of evanescent waves through the LHM slab
due to the amplification of the evanescent waves in the LHM slab. Such a suppression can be effectively
relaxed by reducing the thickness of the LHM slab. We have also tried to eliminate or reduce the destructive influence of the LHM loss to the imaging resolution through manupilating the real and imaginary parts of $\mu_2$ and $\epsilon_2$. However, we found this is not possible (see the appendix).

\acknowledgements
The partial support of Natural Science Foundation of China under a project (60277018) and a key project (90101024) is gratefully acknowledged.

\section*{Appendix: Possibility of reducing the destructive influence of the LHM loss of $\mu_2$ and $\epsilon_2$}
In this appendix, we study the possibility of eliminating or reducing the destructive influence of the LHM loss to the image through manupilating the real and imaginary parts of $\mu_2$ and $\epsilon_2$. Consider the case when the electric permittivity and magnetic permeability of the LHM slab
have the following forms
\begin{eqnarray}
\varepsilon _2  &=&  - \varepsilon _1 (1 - \delta ^{'} ), \nonumber\\
\mu _2  &=&  - \mu _1 (1 - \delta ^{''} ), \nonumber
\end{eqnarray}
where $\delta ^{'}  = \delta _r^{'}  + i\delta _i^{'}$, $\delta ^{''}  = \delta _r^{''}  + i\delta _i^{''}$ ($\delta _r^{'}$, $\delta _i^{'}$, $\delta _r^{''}$, and $\delta _i^{''}$ are all real). We assume that $\delta _i^{'},\delta _i^{''}>0$, and
$|\delta ^{'}|,|\delta ^{''}|\ll 1$. In the following analysis for the LHM slab, our attention will be focused on the evanescent waves for the subwavelength imaging.

From Eq. (3.5), one has
\begin{eqnarray}
\xi  =  - 1 + \frac{1}{2}\left( {\delta ^{''}  + \frac{{\bar \delta }}{{\beta ^2 }}} \right)^2  + O(\bar \delta ^3 ), \nonumber
\end{eqnarray}
where ${\bar \delta }=(\delta ^{'}+\delta ^{''})/2$, $\beta=\sqrt{k_x^2-n_1^2 k_0^2}/(n_1k_0)$ ($k_x > n_1 k_0$ for the evanescent waves). Then the transfer function becomes
\[
{\cal H} \approx \frac{1}{{1 + 0.25(\delta ^{''}  + \bar \delta /\beta )^2 \exp (2D\beta )}}.
\]
Therefore, the requirement for reducing significantly the influence of the LHM loss is
\[
\delta ^{''}=0,\quad \bar\delta=0.
\]
Obviously the above requirement is equivalent to that of $\delta ^{'}=0$ and $\delta ^{''}=0$, which in fact corresponds to the case of an ideal LHM without any loss. Thus, it is impossible to effectively reduce the destructive influence of the LHM loss to the subwavelength imaging by manupilating the real and imaginary parts of $\mu_2$ and $\epsilon_2$.

\begin{figure}
\includegraphics[width=3.5in]{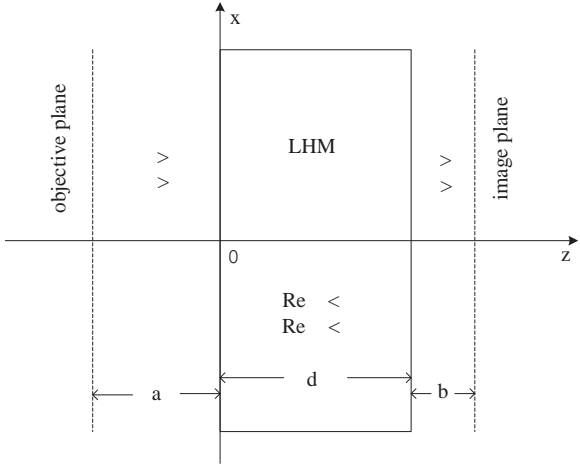}
\caption{Schematic diagram for an imaging system formed by a LHM
slab.}
\end{figure}

\begin{figure}
\includegraphics[width=3.5in]{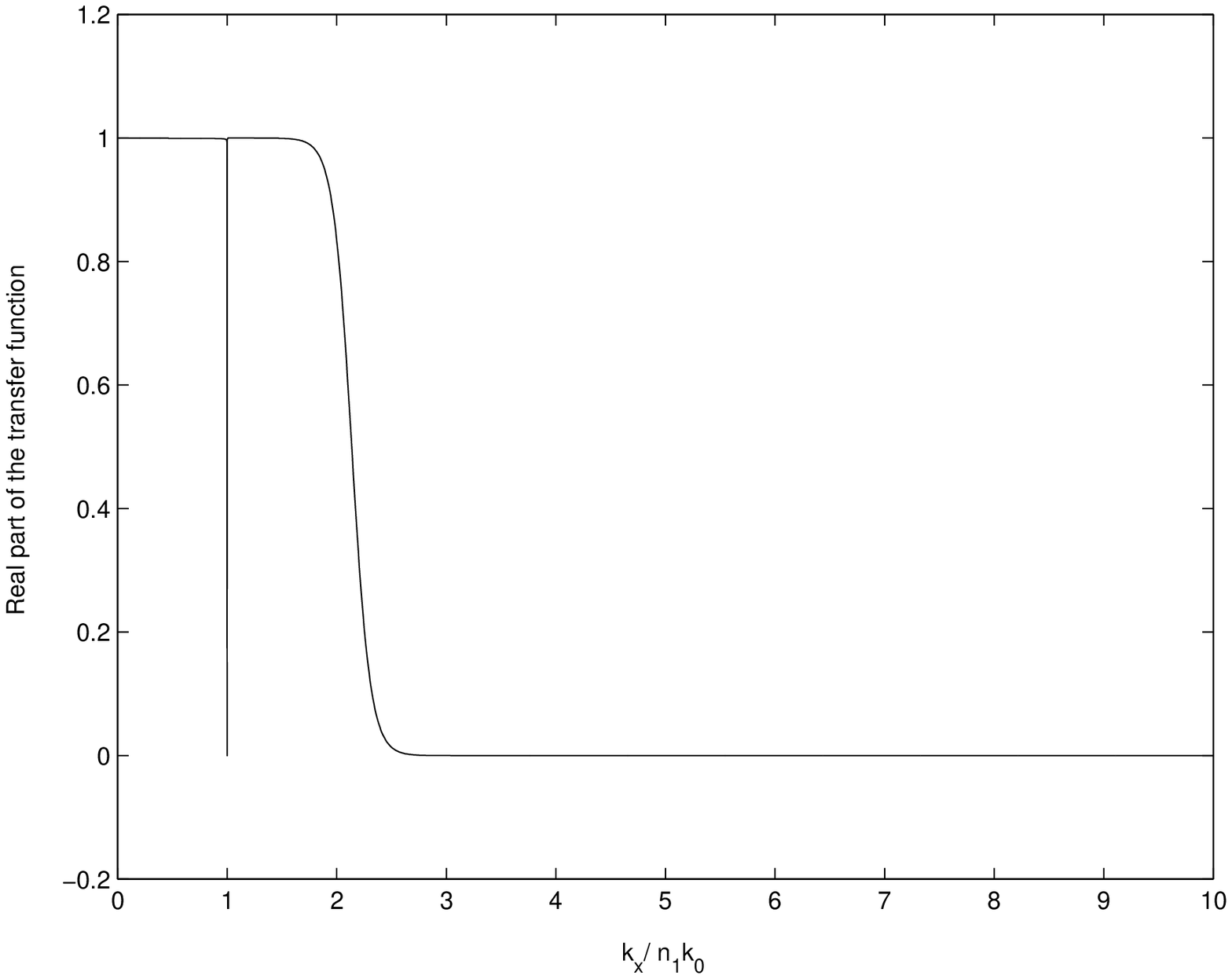}
\includegraphics[width=3.5in]{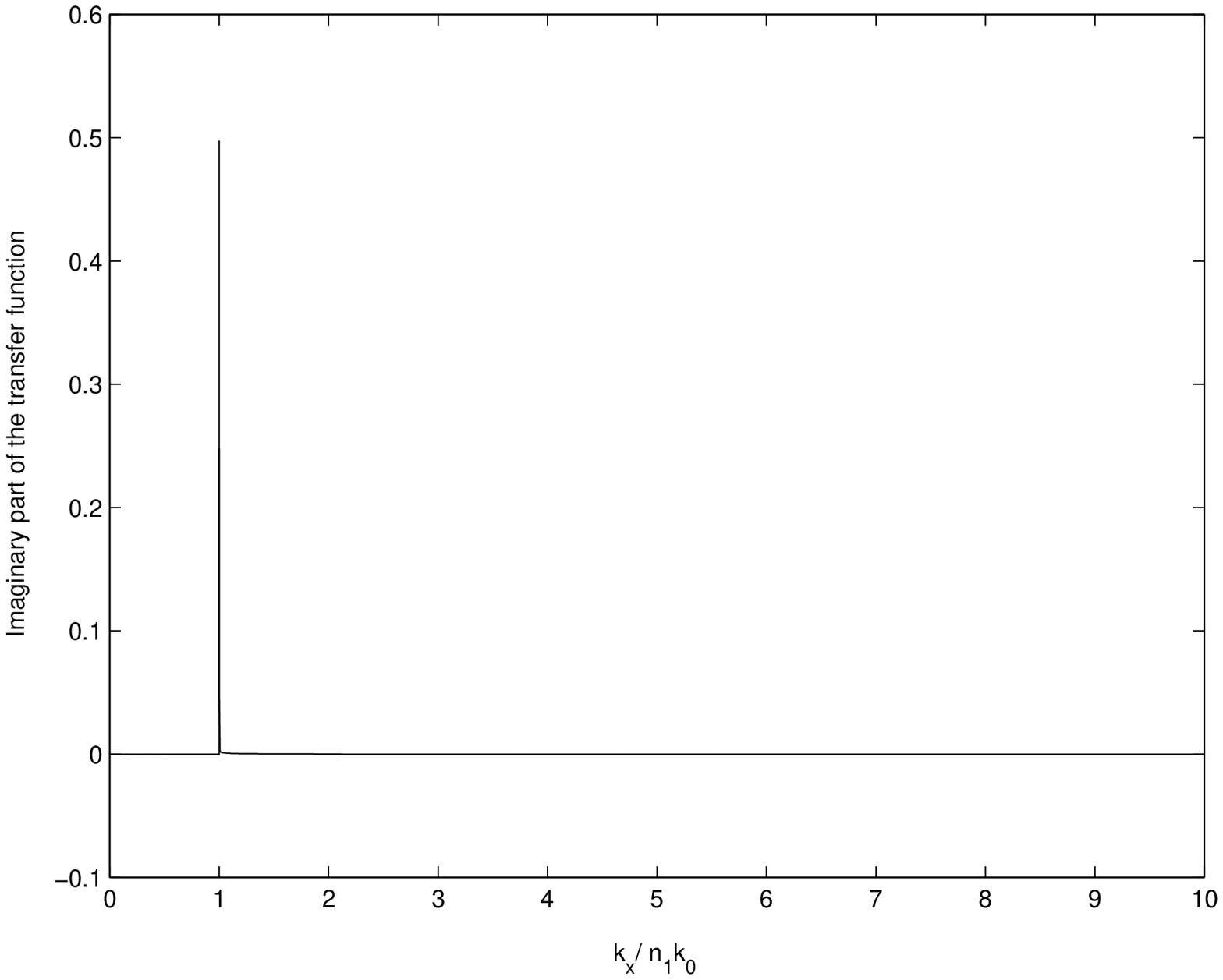}
\includegraphics[width=3.5in]{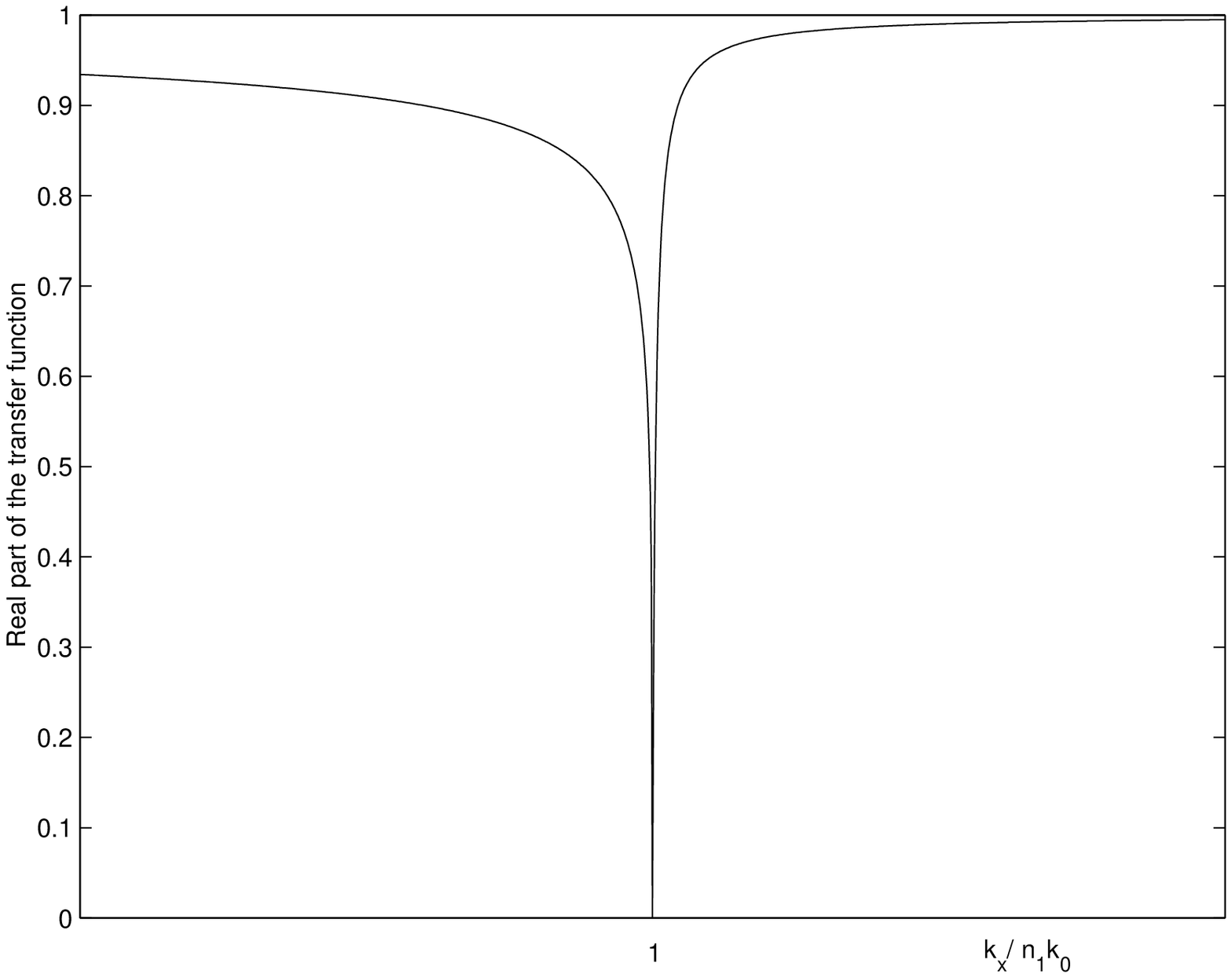}
\includegraphics[width=3.5in]{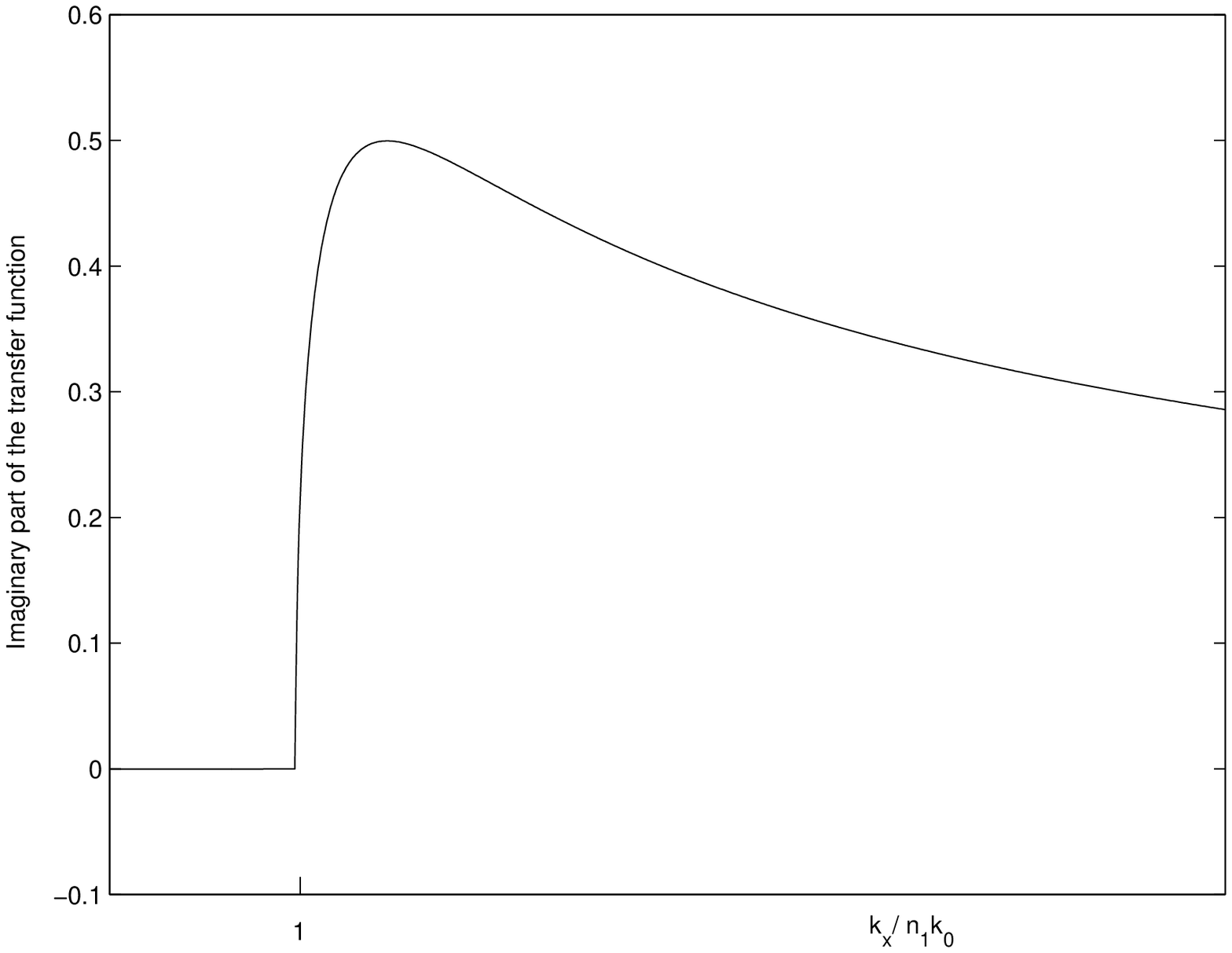}
\caption{(a)The real part  and (b) the imaginary part  of the transfer function ${\cal H}(k_x)$ for  an imaging system of a lossy LHM slab. (c) and (d) give the enlarged views of (a) and (b), respectively, in the neighborhood of $k_x=n_1 k_0$. Here $d=\lambda$ and $\delta=10^{-4}$.}
\end{figure}

\begin{figure}
\includegraphics[width=3.5in]{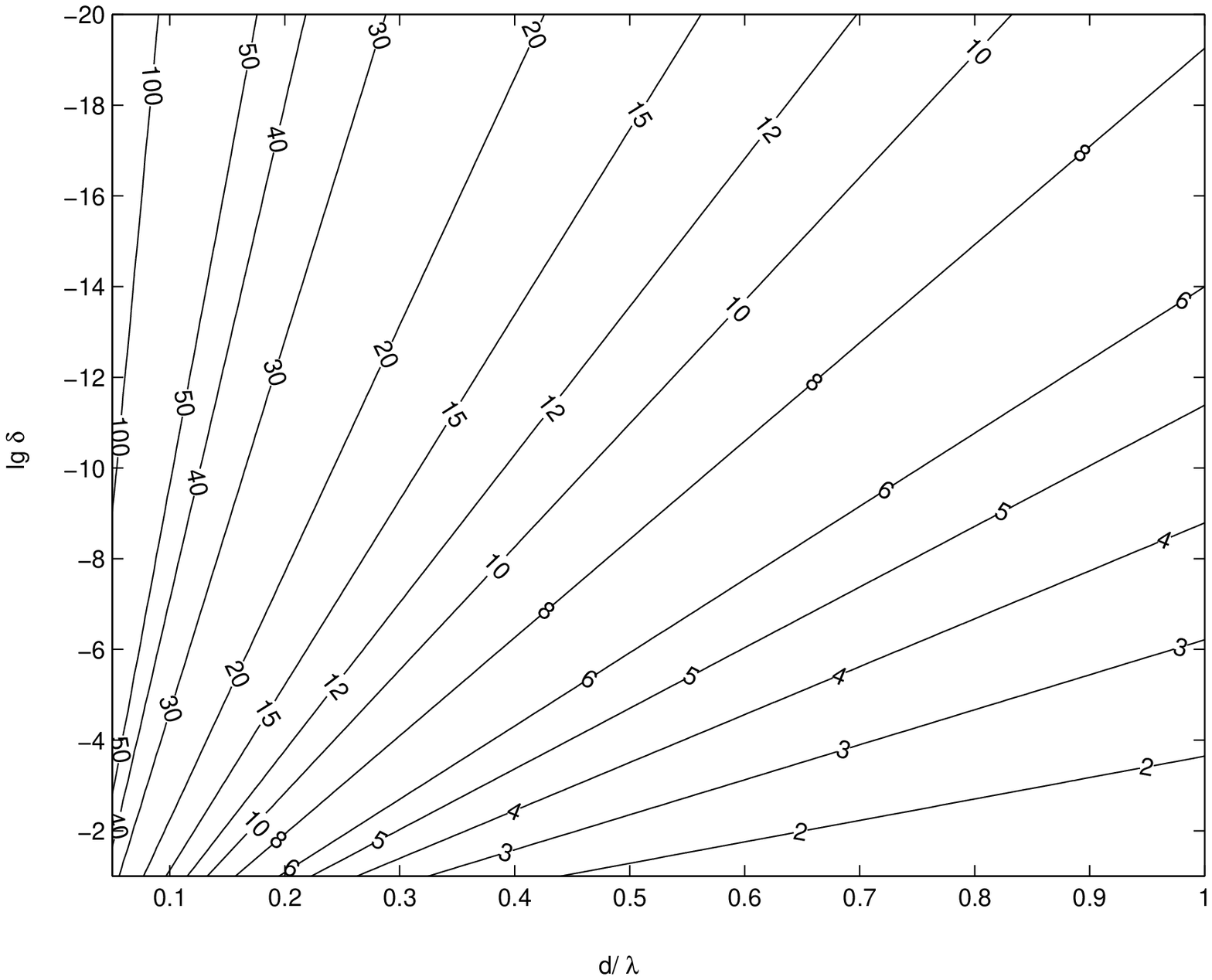}
\includegraphics[width=3.5in]{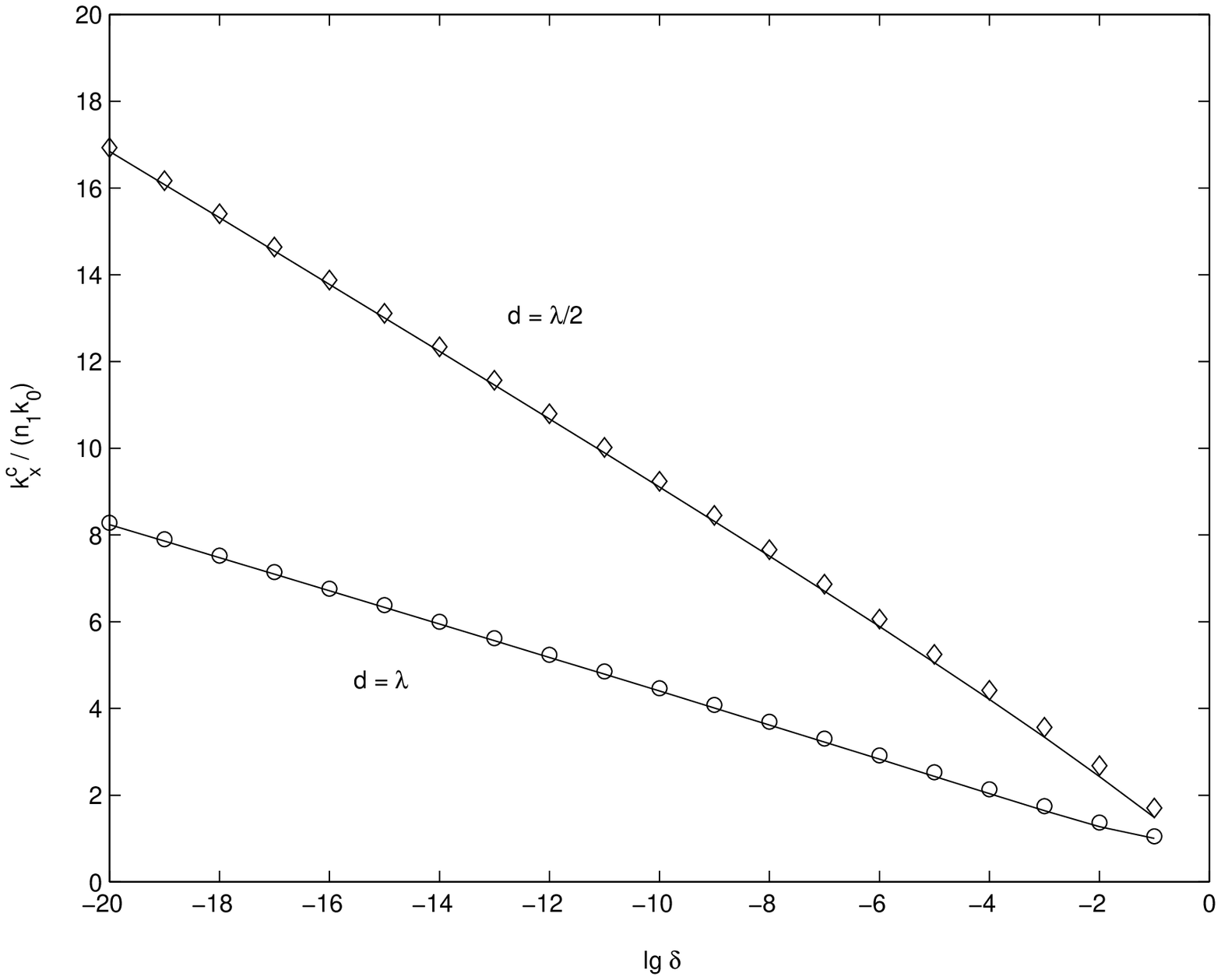}
\caption{(a) Contour plot of $k_x^c/(n_1k_0)$ as a function of the loss parameter $\delta$ and the thickness $d$.
(b) The dependence of the critical $k_x^c/(n_1k_0)$ on the loss parameter  $\delta$ when $d=\lambda$ or $0.5\lambda$. The solid lines
and the marks correspond to the results obtained from Eq. (2) and our analytic formula (8), respectively.}
\end{figure}

\begin{figure}
\includegraphics[width=3.5in]{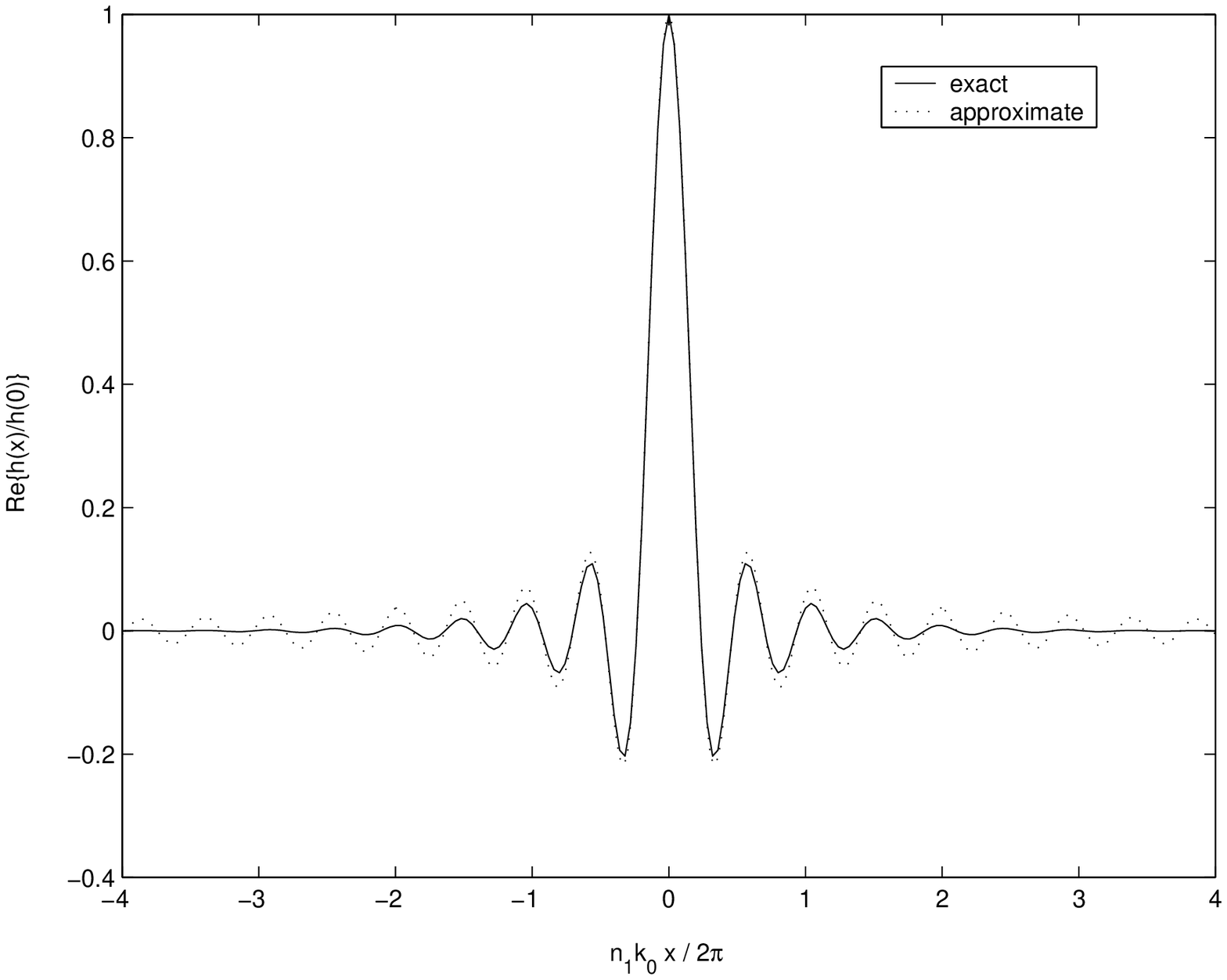}
\includegraphics[width=3.5in]{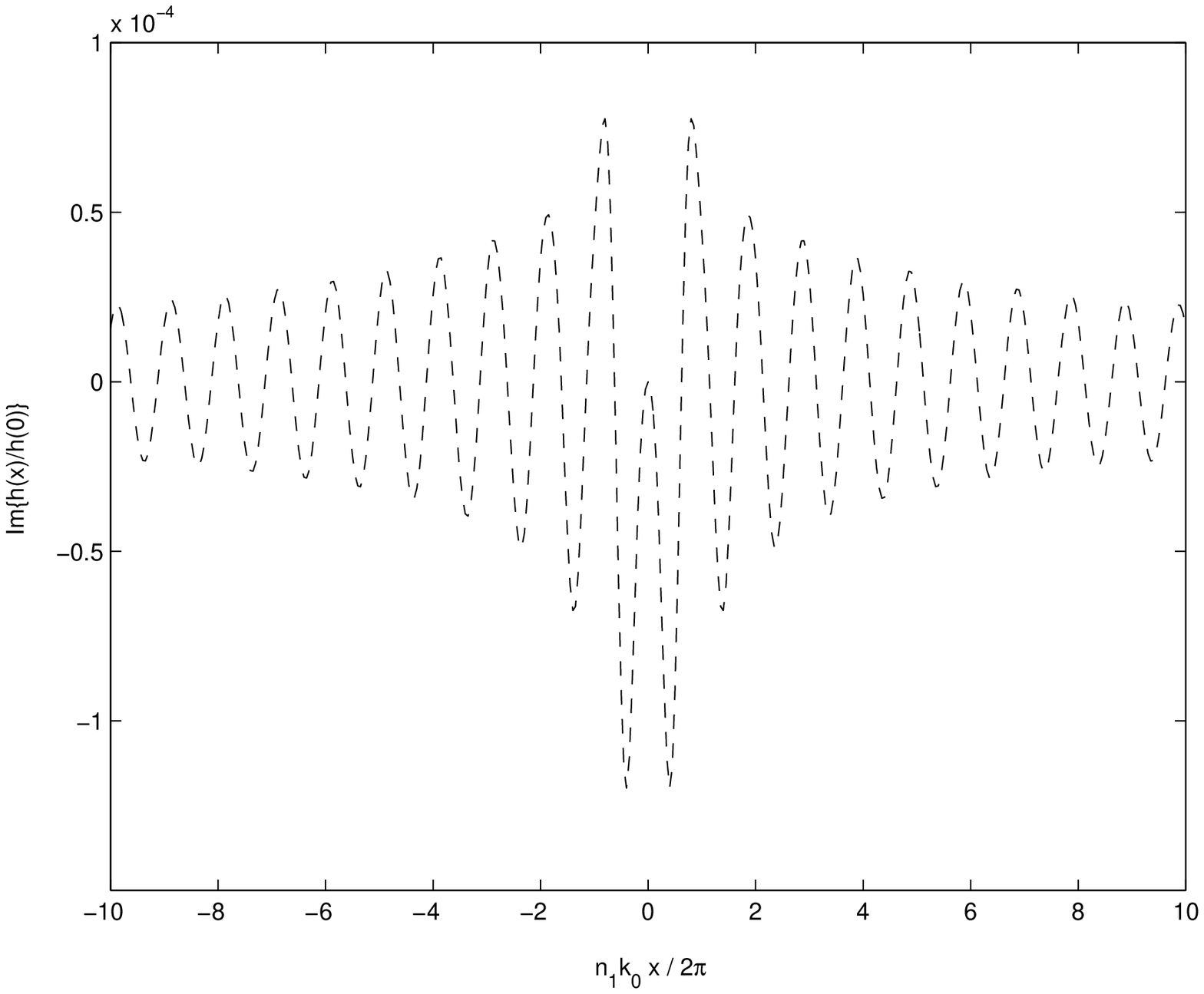}
\caption{(a)The real part of the normalized impulse response calculated from the Fourier transform of Eq. (2) (solid line) and our approximate analytic formula (dotted line). (b)The imaginary part of the normalized impulse response calculated from the Fourier transform of Eq. (2). Here $d=\lambda$ and $\delta=10^{-4}$.}
\end{figure}

\begin{figure}
\includegraphics[width=3.5in]{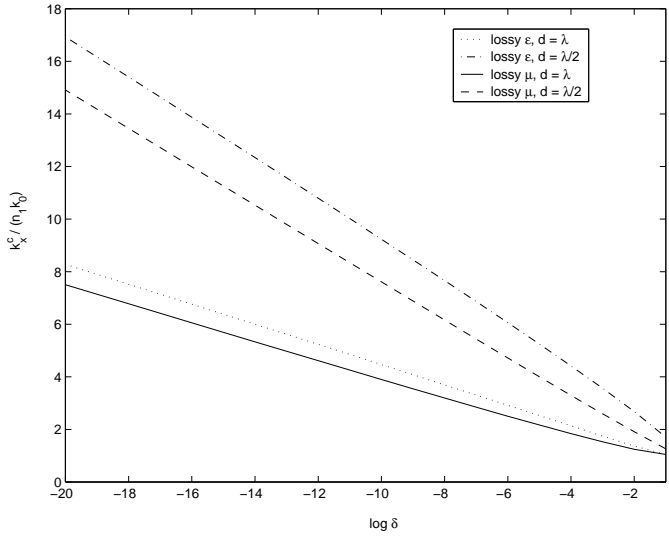}
\caption{Comparison of the critical values of $k_x^c$ when the loss comes from the imaginary
part of $\epsilon$ (solid lines) and the imaginary
part of $\mu$ (dashed lines).}
\end{figure}

\end{document}